\documentclass[twocolumn,showpacs,preprintnumbers,amsmath,amssymb,aps,prl]{revtex4-1}

\usepackage{epsfig}
\usepackage{graphicx,latexsym}

\begin{document}
\title{Geometrical spin manipulation in Dirac flakes\\}
\author{Ioannis Kleftogiannis}
\email{ioannis@ntu.edu.tw}
\altaffiliation{Current address: Department of Physics and Center 
for Theoretical Sciences, National Taiwan University, Taipei 106, Taiwan}
\author{Ilias Amanatidis}
\affiliation{Department of Electrophysics, National Chiao Tung
University,  Hsinchu 30010, Taiwan, Republic of China}

\begin{abstract}
We investigate numerically the spin properties of electrons in flakes made of materials
described by the Dirac equation, at the presence of intrinsic spin-orbit-coupling(SOC).
We show that electrons flowing along the borders of flakes via edge states,
become helically spin-polarized for strong SOC,
for materials with and without a gap at the Fermi energy,
corresponding to the massive and massless Dirac equation
respectively. The helically spin-polarized electrons
cause geometrical spin splitting on opposite sides of the
flakes, leading to spin-resolved transport controlled by the flake's geometry
in a multi-terminal device setup. A simple analytical model containing the basic ingredients of the problem
is introduced to get an insight of the helical mechanism, along with
our numerical results which are based on an effective tight-binding model.
\end{abstract}


\maketitle
\section{Introduction}

Mesoscopic low-dimensional materials described by the Dirac equation for relativistic
particles is a rapid developing field providing a  vast area of fundamental
research and promising a wide range of applications. One of the earliest examples is
graphene, a two-dimensional (2D) sheet of carbon atoms arranged in a honeycomb
lattice structure resulting in an electronic behavior at the Fermi level that resembles
relativistic massless particles described by the massless Dirac equation \cite{novoselov2004,Neto}.
Another more recent example is transition metal dichalcogenide
(TMD) monolayers\cite{Frindt,Splendiani,Radisavljevic,Heine,Ataca,Gomez,Taniguchi1,Yuan,Lee},
with honeycomb lattice structure similar to graphene but with a varying gap at the Fermi level, for instance
1.35~eV for W${\rm S}_2$ and 1.83~eV
for Mo${\rm S}_2$. Alternatively a considerable
gap of up to 0.278~eV can be created in graphene by embedding it
on boron nitride (BN) substrates\cite{Menno,Moon,Chizhova}.
Graphene on BN substrates and TMD monolayers can be effectively described
by the massive Dirac equation, instead of the massless one which describes pristine
graphene.

Dirac materials can be easily formed in confined dot-like
structures of various sizes and shapes, known as flakes,
having been studied theoretically and demonstrated experimentally
\cite{novoselov2004, yamamoto,ezawa,wang,heiskanen1,heiskanen2,
ponomarenko,wu,zu,Schnez,Pavlovic,Chhowalla,Zhang}.
Specifically for TMD flakes the trigonal shape
is a natural state in their experimental preparation\cite{Chhowalla,Zhang}.
The confinement in Dirac flakes leads to
unconventional electronic properties such as the edge states,
whose electronic density is concentrated at the edges
of the flakes \cite{fujita,ezawa,wang,heiskanen1}
leading to novel topological phenomena
\cite{wakaplus,akhemerov,kleftogiannis,amanatidis}.
Edge states are fundamentally a consequence of the honeycomb
lattice morphology present in Dirac materials
that favors destructive interference effects
when zig-zag edges are present.

On the other hand spin-orbit-coupling(SOC) in Dirac materials offers
the possibility to manipulate the electron spin for integration
in spintronic applications \cite{Loss1998,Zutic2004,Wolf2001,Winkler2003}.
Despite the weak SOC in graphene, the interaction
with the system's confinement, particularly in graphene nanoribbons,
has been shown to produce novel quantum phases like 
the Quantum-Spin-Hall(QSH) effect\cite{Kane,Konig,Chen}.
The QSH effect has sparked a continuously increasing interest
in exotic materials that display related behaviors known as topological insulators.
Also spin-transport effects have been investigated in graphene 
flakes via DFT simulation\cite{Ciraci,Ono}.
In the same sense, TMD monolayers and other emergent 2D materials 
are also promising candidates for relevant theoretical
investigations\cite{Guinea} and applications since the SOC strength
can be much higher than graphene, leading to large spin-spilittings
up to 0.46~eV in some cases of TMD monolayers\cite{Zhu}.

In the current work we analyze the spin properties of electrons
in Dirac flakes with intrinsic SOC via the numerical calculation of the transmission probabilities
along the border of the flakes. We show that for sufficiently strong
SOC the electrons that propagate in opposite directions along the
flake's border via the edge states, have opposite spins, leading
to helically spin-polarized electrons. 
We demonstrate this property in a multi-terminal device setup naturally formed 
by attaching three leads at the corners of a trigonal flake. Appropriate tuning
of the chemical potential of the leads
creates spin-resolved electronic transport with opposite spin-polarization
along the opposite sides of the triangle.
This allows a geometrical manipulation of the spin intrinsically without the need to apply
external fields. The helically spin-polarized electrons are present
in materials with and without a gap at the Fermi
energy, described by the massive and massless Dirac equation, respectively,
which could correspond to TMD monolayers and graphene. Additionally, we introduce a
simplified analytical model, that contains the fundamental ingredients
of the numerical model, allowing us to get a better understanding of our numerical
results.

\section{Model}

\begin{figure}
 \includegraphics[width=\columnwidth,angle=0,scale=1.0,clip=true]{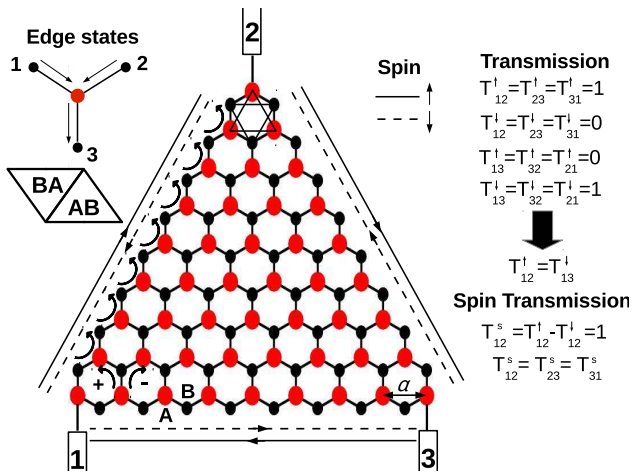}
 \caption{A trigonal Dirac flake with the characteristic
 honeycomb lattice structure consisting of two sublattices A and B.
 An example of the spin-orbit coupling interaction involving spin-dependent
 hopping amplitude between second-nearest neighbor sites can be seen inside
 a single hexagon.
 Three leads can be attached naturally to the corners of the trigonal flake.
 The arrows across the flake's border indicate the helically spin-polarized electronic
 flow, satisfying the relations for the transmission probabilities between
 different terminals, shown on the right.
 We characterize the flake size by the length L of the triangle base,
 in units of the lattice constant $a$. The edge state mechanism due to
 the honeycomb lattice structure can be seen on the left along
 with a schematic showing how to exchange the A and B atoms by cutting the flake
 in different orientations.}
  \label{fig1}
 \end{figure}

In this section, we present our numerical tight-binding model to
simulate flakes made out of Dirac materials at the presence of SOC.
Dirac materials can be described by the massive Dirac
Hamiltonian\cite{Guinea,dloss,Andor,Levitov,Xu,Ioannis}
\begin{equation}
{\cal H}_{0} = \hbar\upsilon_f(k_x \tau_z \sigma_1 + k_y\sigma_2) + V\sigma_3
\label{h_0}
\end{equation}
where $\upsilon_f$ is the Fermi velocity of electrons whose value depends on
the material under investigation, and $\sigma_i$ being the Pauli
matrices acting on orbital space. Symbol $\tau_z=\pm1$ denotes the non-equivalent valleys
that are present in graphene and other Dirac materials like TMD monolayers, at the six corners of their hexagonal
Brillouin zone. Eq. \ref{h_0}
describes relativistic particles of mass $V$, while the speed of
light $c$ is replaced by $\upsilon_f$. Different values of $V$
classify different materials for example $V=0$~meV and $\upsilon_f
\approx 10^6$~m/s corresponds to graphene, while finite $V$ could
describe TMD monolayers and graphene on BN
substrates\cite{Chizhova,dloss,Guinea,Andor}.

In order to perform our numerical calculations, we
consider an effective tight-binding model, consisting of a honeycomb
lattice with first nearest neighbor hopping, along with an onsite staggered potential
simulating the mass term in Eq. \ref{h_0},
\begin{equation}
\label{h_0_tb}
H_0= \sum_{i}\epsilon_{i}c_{i\mu}^\dagger c_{i\mu}+
\sum_{\langle i,j\rangle}t_{i,j} (c_{i\mu}^\dagger c_{j\mu}
+c_{j\mu}^\dagger c_{i\mu} )\, ,
\end{equation}
where $c_{i\mu}^\dagger$ ($c_{i\mu}$) is the creation (annihilation)
operator for spin $\mu$ at site $i$ while $t_{i,j}=1$~eV is a uniform
hopping between all the nearest neighbor lattice sites.
The onsite potential $\epsilon_i$ is $V$ and $-V$
on $A$ and $B$ sublattice sites respectively with $V=0$
corresponding to graphene. The staggered potential breaks the
inversion symmetry resulting in a gap $2V$ at the Fermi energy for
infinite unbounded systems. At low
energies the effective tight-binding Hamiltonian Eq. \ref{h_0_tb} transforms
to Eq. \ref{h_0} with $\hbar\upsilon_f=\frac{t\sqrt{3}a}{2}$
where $a$ is the lattice constant. This effective tight-binding model
can be thought as a numerical version of the massive Dirac equation.

The SOC can be introduced in the tight-binding model by considering
spin-dependent hoppings along next nearest neighbors in the honeycomb lattice as\cite{dloss,Kane},
\begin{equation}
\label{h_soc}
  H_{SOC}=\alpha \sum_{\langle\langle i,j\rangle\rangle,\mu,\mu^{'}}\nu_{i,j}c_{i\mu}^\dagger s_{z,\mu,\mu^{'}}c_{j\mu^{'}},
\end{equation}
where the sum runs over next nearest neighbors on all the lattice sites. The spin-dependent amplitude is $\nu_{i,j}=1(-1)$
when the electron at site $i$ makes a anticlockwise(clockwise) 
turn in order to hop from the first nearest to the second nearest neighbor $j$, represented by 
the curved arrows inside the honeycomb lattice of the trigonal flake in Fig. \ref{fig1}.
The strength of the intrinsic SOC is determined by $\alpha$. We note that the SOC interaction described by Eq. \ref{h_soc}
involves connections between sites belonging to the same sublattices A or B, in conjunction
with the first nearest neighbor term in Eq. $\ref{h_0_tb}$ which involves connections only between A and B sublattice sites.
Eq. $\ref{h_soc}$ is responsible for the QSH effect in confined graphene systems,
acting as a bridge between graphene and topological insulators.
The total Hamiltonian of our system is
\begin{equation}
\label{h_total}
  H=H_0+H_{SOC}.
\end{equation}
By applying hard-wall boundary conditions on Eq. \ref{h_total} we can simulate
the trigonal flakes, like the one shown in Fig. \ref{fig1}.

In order to calculate the transmission probabilities via the edges
of the trigonal flake, we attach perfect semi-infinite linear chains described by
$H_{1d}= \sum t(c_{i\mu}^\dagger c_{i+1\mu}
+c_{i+1\mu}^\dagger c_{i\mu} )$, with hopping t=1eV
at it's three corners as shown in Fig. \ref{fig1},
forming this way a three terminal device. The energy E, 
the SOC strength $\alpha$ and V, are all reported in units of t. We observe that
Eq. \ref{h_soc} contains only $s_z$ so that the total
Hamiltonian of the system Eq. \ref{h_total} can be split in spin up and down diagonal blocks,
resulting in zero transmission probability between opposite spins(see Appendix).
Therefore the transmission probability between different terminals can be written as $T^{\mu}_{ij}$
where i,j=1,2,3 runs over the terminals and $\mu=\uparrow,\downarrow$
denotes the spin orientation of the electron.
We define also the spin transmission as $T^{s}_{ij}=T^{\uparrow}_{ij}-T^{\downarrow}_{ij}$
which can be used to estimate the degree of spin-polarization.
$T^{\mu}_{ij}$ can be calculated by using the Green's function
${\cal G}(i,j,\mu,E)=(E^{+}-H-\Sigma_1(E^{+})-\Sigma_2(E^{+})-\Sigma_3(E^{+}))^{-1}$ with $E^{+}=E+i\eta$, where E is the incident
energy and $\eta \rightarrow 0^{+}$, while $\Sigma_i(E)$ is the
self energy of each semi-infinite chain used as leads, being diagonal
in the spin-basis, since we neglect the SOC inside the leads(see Appendix).
By using the velocities inside the leads according to the Fisher-Lee relations we get
$T^{\mu}_{ij}=(4-E^{2})\vert {\cal G}(j,i,\mu,E) \vert^2 $.

%
%

\section{Edge states}

\begin{figure}
\begin{center}$
\begin{array}{cc}

\includegraphics[width=0.5\columnwidth,clip=true]{fig2a.eps}
\includegraphics[width=0.5\columnwidth,clip=true]{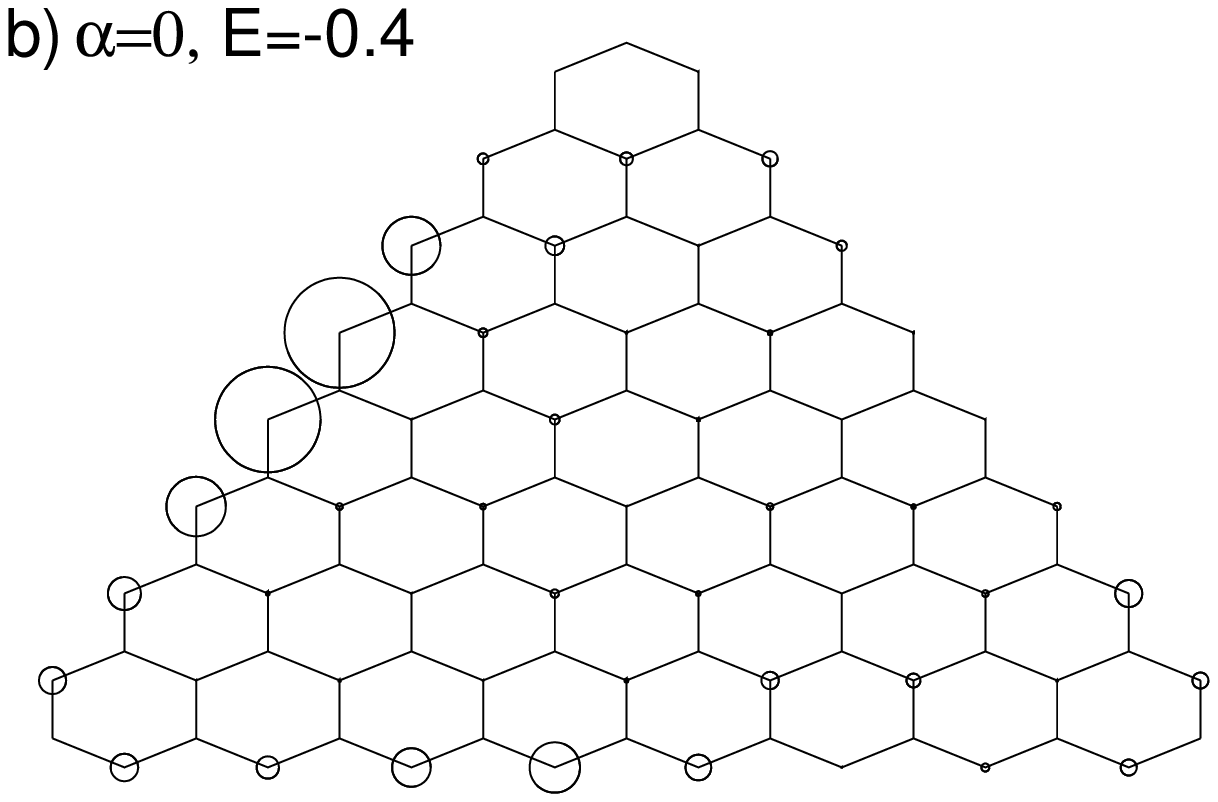} \\

\includegraphics[width=0.5\columnwidth,clip=true]{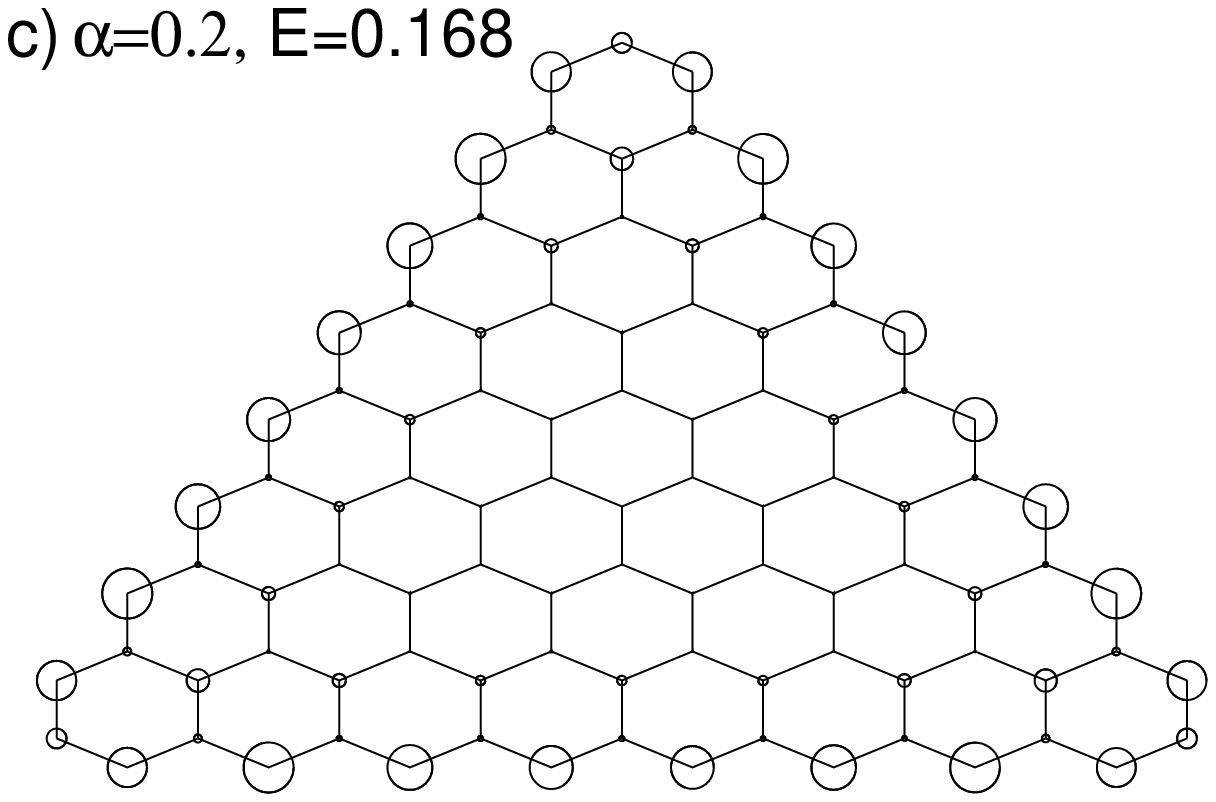}
\includegraphics[width=0.5\columnwidth,clip=true]{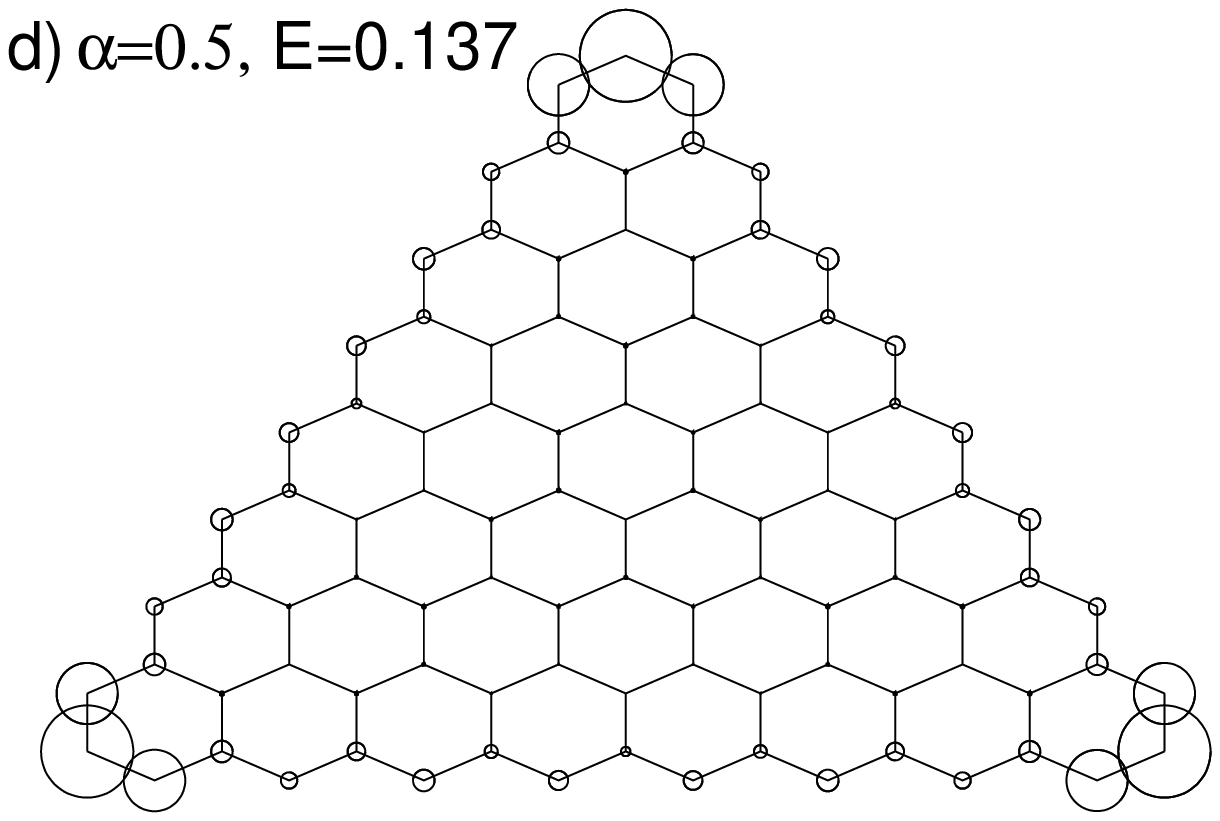}

\end{array}$
\end{center}
\caption{a) The energy levels of a trangular Dirac flake for L=8, V=0.4 and different SOC strengths $\alpha$ near the gap induced by V.
Fourteen edge states at E=-V can be  distinguished for $\alpha=0$ due to sublattice symmetry, dispersing inside the gap for finite $\alpha$.
The levels become almost homogeneously distributed for large $\alpha$. b) The respective wavefunction for $\alpha=0$ at E=-V is an edge state
being non-zero only on B sublattice sites. c) The edge state is enchanched by the SOC becoming completely concentrated at the edges.
d) The wavefunction is accumulated at the corners for large $\alpha$.}
\label{spectrum}
\end{figure}

In this section, we analyze the edge states present in the trigonal zig-zag flakes
made of Dirac materials, which are responsible for the electronic behavior near the Fermi level.
In the lattice representation, the number of edge states
can be derived by extending chiral symmetry arguments applied on the honeycomb lattice which is bipartite\cite{Inui}.
In general the Dirac flakes can be split into two sublattices with
$N_A$ number of A and $N_B$ number of B sublattice sites, as shown in Fig. \ref{fig1}.
The type of the outermost atoms
at the zig-zag edges of the flake determines which number, either $N_A$ or $N_B$ is larger.
For the configuration in Fig. \ref{fig1} with B type atoms, we have $N_B>N_A$,
while $N_B-N_A=L-1$, with L being the length of the triangle basis in units of the lattice
constant $a$.
We note that the type of atoms A and B can be easily exchanged by changing the orientation of
the triangle base on an infinite graphene sheet by 60 degrees, compared to the base of the flake
in Fig. \ref{fig1}, resulting in $N_A>N_B$ in this case.
Since the spin-orbit coupling given by Eq. \ref{h_soc} contains only $s_z$ it does not mix
the spin up with the spin down orientations, so that the total
Hamiltonian of the system can always be split into spin up and spin down diagonal blocks.
Consequently we can write down the Schr\"{o}dinger difference equations centered
on A and B atoms for spin up(down) denoted by $\mu=1(-1)$ as,

\begin{equation}
\label{schr_chiral}
\begin{array}{cc}

(E-\epsilon_A)\Psi_{A,i}^{\mu}= t\sum_{j}\Psi_{B,j}^{\mu}+
i\mu\alpha\sum_ j \nu_{i,j} \Psi_{A,j}^{\mu} \\ \\

(E-\epsilon_B)\Psi_{B,i}^{\mu}= t\sum_{j}\Psi_{A,j}^{\mu}+
i\mu\alpha\sum_ j \nu_{i,j} \Psi_{B,j}^{\mu}

\end{array}
\end{equation}
where $\Psi_{A(B),i}$ is the wavefunction amplitude on site i of each sublattice and $\epsilon_{A(B)}$ is the
staggered on-site potential. In both sets of equations, the first
term on the right-hand side comes from the nearest neighbour hopping,
while the second term is associated with the
second nearest neighbor hopping with spin dependent amplitude $\nu_{i,j}$,
as in Eq. \ref{h_soc}. In the absence of SOC ($\alpha=0$),
we can easily derive the number of edge states by algebraic arguments
applied on Eq. \ref{schr_chiral}.
For $\epsilon_A=V,\epsilon_B=-V$ the equations become,
\begin{subequations}
\renewcommand{\theequation}{\theparentequation.\arabic{equation}}
\begin{align}
(E-V)\Psi_{A,i}^{\mu}= t\sum_{j}\Psi_{B,j}^{\mu}\label{schr_chiral_nosoc_1} \\
(E+V)\Psi_{B,i}^{\mu}= t\sum_{j}\Psi_{A,j}^{\mu}.\label{schr_chiral_nosoc_2}
\end{align}
\end{subequations}

Eq. \ref{schr_chiral_nosoc_1} is a set of $N_A$ equations with $N_B$ unknowns written on the A sites,
while  Eq. \ref{schr_chiral_nosoc_2} is a set of $N_B$ equations with $N_A$ unknowns written on the B sites, respectively.
We observe that for E=-V, Eq. \ref{schr_chiral_nosoc_2}
transforms to the homogeneous set $0= t\sum_{k}\Psi_{A,k}^{\mu}$
which can only be satisfied by assuming that $\Psi_A^{\mu}=0$
since there are more equations than unknowns.
On the other hand Eq. \ref{schr_chiral_nosoc_1} transforms to
$-2V\Psi_A^{\mu}= t\sum_{k}\Psi_{B,k}^{\mu}\Rightarrow0=t\sum_{k}\Psi_{B,k}^{\mu}$ which gives $\Psi_B^{\mu}$,
but results also in $N_B-N_A$ linearly independent solutions, since there are less equations
than unknowns.

Consequently for $\alpha=0$ there are at least $N_B-N_A$ states at E=-V
with non-zero amplitudes only on the B sublattice sites.
For V=0 we derive the well-known case of bipartite lattices
where there are at least $N_B-N_A$ states at E=0 due to the chiral symmetry\cite{Inui}.
With the application of the staggered potential ($V\neq0$) the energy of these states is shifted to E=-V.
Considering the spin these states become doubly degenerate.

Due to the destructive interference
effects favored by the honeycomb lattice when zig-zag edges are present,
these states either at E=-V or E=0 are edge states with their corresponding
wavefunctions concentrated at zig-zag edges of the flakes.
The edge state mechanism can be robustly understood by isolating
part of the zig-zag edge as shown in Fig. \ref{fig1}
at the right side of the flake schematic, where there are three B sites surrounding one A site.
The edge states can be visualized as two incoming electronic
waves from two B sites 1 and 2, interfering and
giving an outgoing wave at site 3 site via the A site.
For E=0 there is destructive interference of the two incoming waves due
to a phase difference between the wavefunction amplitudes on the respective sites 1 and 2 , 
introduced by the Bloch's theorem, as 
can be seen easily in a semi-infinite graphene sheet with one zig-zag edge\cite{fujita}. 
These two waves from sites 1 and 2 interefere destructively
via the central A site whose wavefunction is zero, 
giving zero amplitude at site 3 for states at the ends of the Brillouin zone where
the phase difference is $\pi$ or a reduced amplitude otherwise. This intereference process continues 
inside the rest of the honeycomb lattice and is present whenever zig-zag edges exist
at the borders of a confined graphene structure.
The application of the staggered potential, does not alter this basic topology of
the honeycomb lattice structure which is responsible for the edge states,
since it gives a potential -V on all the B sites 1,2,3 and potential V on site A,
resulting in the shifting of the edge state mechanism to energy E=-V.
We note that the intereference mechanism responsible for the edge states 
vanishes for the armchair edge morphology.

According to the above algebraic arguments, the energy of the edge states can
be shifted at the opposite (conduction) side of the energy spectrum at E=V by simply exchanging
the potential on A and B sublattice sites,
so that the outermost sites at the zig-zag edges in Fig. \ref{fig1} 
have on-site potential V instead of -V.
We summarize that for Dirac flakes with zig-zag edges, in the absence of SOC,
there are at least $\mid N_{A}-N_{B} \mid$ edge states at E=V(E=-V)
when the potential of the outermost atoms of the zig-zag
edges is V(-V). Therefore, there are always edge states whose energy is
determined by the type of atoms at the zig-zag edges of the flake, while their
number increases linearly with the system size L as $\mid N_{A}-N_{B} \mid=L-1$ .
In realistic systems, our results imply that the energy of the edge states can be tuned
near either the valence or the conduction band edge,
by appropriately cutting the Dirac flakes, in such a way that the type of atoms
A and B are exchanged.

When the SOC is introduced in the system, for $\alpha\neq0$ in Eq. \ref{schr_chiral}, the edge states disperse inside
the gap created either, due the staggered potential V, or due to the finite flake size.
For instance, this phenomenon has been observed in Mo${\rm S}_2$ flakes by
multi-orbital tight-binding simulation\cite{Pavlovic}. In our case this behavior can be understood by
the SOC term in Eq. \ref{schr_chiral} which alters
the values of E for which these equations can be satisfied according
to the algebraic arguments we presented. The energy E=-V is no longer a solution of these equations,
while the wavefunction amplitude becomes gradually finite on both sublattice sites A and B as $\alpha$ is increased.
In the simulation of specific materials the SOC term Eq. \ref{h_soc} has to be applied only on one of the sublattices.
If it is applied only on A then the results we derived for $\alpha=0$ remain valid,
since the wavefunction amplitude on A sites is zero and any perturbation
applied on this sublattice will not affect the system's properties.

We have verified our analytical results numerically. An example
can be seen in Fig. \ref{spectrum}a where the energy levels
for a trigonal flake with L=8 and V=0.4 are shown, near the energy gap.
All states come in pairs due to the spin degeneracy, which is preserved even for for finite $\alpha$.
In agreement with our algebraic arguments there are $2(N_{B}-N_{A})=14$ edge states at
E=-V, which disperse inside the gap for finite $\alpha$.
For sufficiently large $\alpha$ the states tend to repulse, becoming almost
homogeneously distributed inside the whole gap,
despite being initially concentrated at it's lower end(E=-V).
In general a significant interplay between the SOC and the gap generated
by the staggered potential is expected.
Additionally the edge states due to zig-zag edges become gradually
mixed with the edge states created due to the SOC in analogy to the edge state mechanism in the Quantum-Hall effect.
In this sense, the edge state mechanism originating from the honeycomb
lattice structure is enhanced by the SOC,
however it's contribution becomes irrelevant for strong SOC. 

Some examples of these edge states can be
seen in Fig. \ref{spectrum}b,c,d where the wavefunction propability is shown, 
represented by the radius of circles at each site. For $\alpha=0$ the amplitude of the edge
state at E=-0.4, is non-zero only on the B sites, with the largest amplitude residing along the edges.
There is almost perfect concentration of the wavefunction
at the zig-zag edges of the flakes for finite $\alpha$. A
gradual accumulation at the corners of the triangle can be observed
for large SOC ($\alpha=0.5$). This is due to the
different lattice connectivity between the sites at the corners
of the triangle and the ones along the zig-zag
edge. Particularly the corner sites
are connected via two spin-dependent hoppings to second
nearest neighbors, instead of four
for sites along the zig-zag edge, coming from Eq. \ref{h_soc}. Also, each of the corners
can be thought as the intersection between two zig-zag chains causing
additional intereference effects, that cannot be distinguished
for $\alpha=0$ since the wavefunction probability
is zero on A sites. We note that the results are similar for the spin up and down cases.

\section{Transmission probabilities}

\begin{figure}
\includegraphics[width=\columnwidth,scale=1,clip=true]{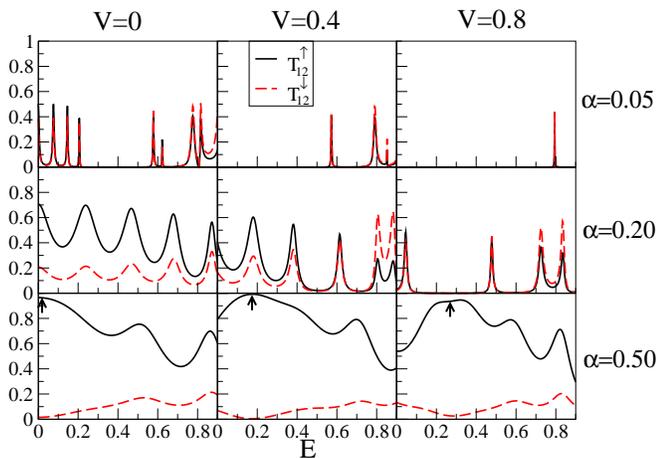}
\caption{The transmission probabilities $T^{\mu}_{12}$  between terminals 1 and 2 for
spin up and down, different $\alpha$ and V. For strong SOC
$\alpha=0.5$ a significant difference between up and down is formed indicated by an arrow
where $T^{\uparrow}_{12}\approx1$ and $T^{\downarrow}_{12}\approx0$,
signifying helically fully spin-polarized electrons flowing along the edge.
The helical regime is shifted along the energy for increasing V.}
\label{tupdown}
\end{figure}

\begin{figure}
\begin{center}$
\begin{array}{cc}

\includegraphics[width=1\columnwidth,clip=true]{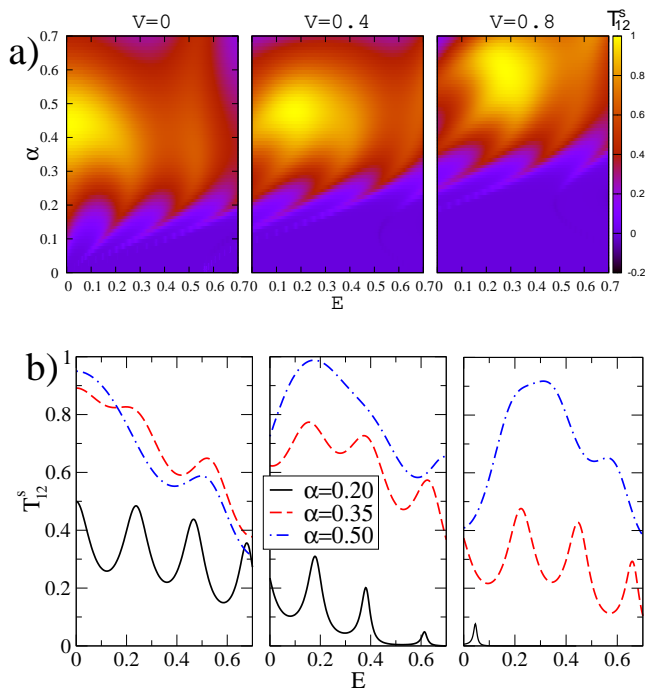} \\
\includegraphics[width=0.97\columnwidth,clip=true]{fig4b.eps}

\end{array}$
\caption{a) Density color plot of the spin transmission versus E and $\alpha$.
Inside the yellow area where $T^{s}_{12}\approx1$ the electrons are helically
spin polarized. Larger $\alpha$ is required for materials
with wider gaps(larger V)in order to observe the helicity.
b) Some respective cases of $T^{s}_{12}$ versus E for different $\alpha$.}
\label{tsphase}
\end{center}
\end{figure}

\begin{figure}
\includegraphics[width=1\columnwidth,clip=true]{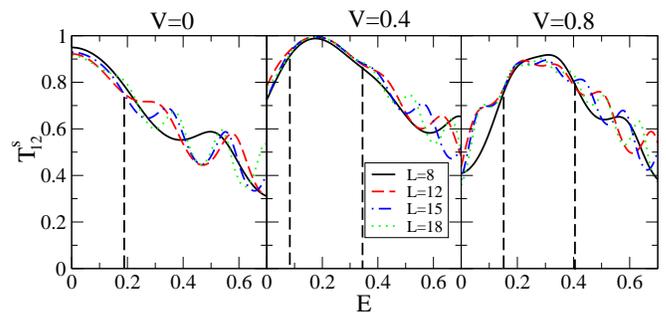}
\caption{$T^{s}_{12}$ for $\alpha=0.5$ and different flake sizes L.
The helical regime is retained for all L independently of V.}
\label{tssize}
\end{figure}

In the current section we investigate the transmission probabilities via the edges of the trigonal
Dirac flakes with SOC. In order to probe the transmission we form a three terminal
device setup by attaching linear chains at the three corners of the triangle,
as shown in Fig. \ref{fig1}. In Fig. \ref{tupdown} we show the spin-dependent transmission probabilities $T^{\mu}_{12}$
for L=8 along one side of the triangle from terminal 1 to 2 for the spin-up and spin-down cases.
We have performed the calculation for different values of the
staggered potential V and the SOC strength $\alpha$. The electronic flow inside the energy area where we plot our results,
is carried via edge states like the ones in Fig. \ref{spectrum} for finite $\alpha$.
For weak SOC ($\alpha=0.05$) resonant transmission via the respective energy levels of the system can be seen
as sharp peaks. Increasing $\alpha$ causes broadening of these peaks resulting in the smooth fluctuations shown.
Clearly, a significant difference between the opposite spins is starting to show for sufficiently large $\alpha$.
For $\alpha=0.5$ a small regime is formed for every V, indicated with a small arrow,
where there is almost perfect propagation of solely spin-up electrons
($T^{\uparrow}_{12}\approx 1,T^{\downarrow}_{12}\approx 0$).
We note that large SOC interactions have been demonstrated in  
TMD monolayers\cite{Zhu} which correspond to finite V in our model. 

This behavior can be attributed to two mechanisms. 
One is the helical currents at the edges induced by the QSH effect
followed by the Laughlin's argument, 
in analogy with the zig-zag nanoribbons with SOC\cite{Kane}. An additional mechanism can be
revealed by a carefull observation of the zig-zag edge. In Fig. \ref{fig1}
we can see that the electronic flow from terminal 1 to 2 is carried through the
outermost B(black) sites via anticlockwise turns 
corresponding to positive spin-flip amplitude $\nu=+1$
in Eq. \ref{h_soc}, implying spin-up polarization. The other possible
flow via A(red) sites across the edge corresponding to spin-down polarization
is partially supressed since the wavefunction amplitude is generally smaller than on B sites,
being a remnant of the chiral symmetry for $\alpha=0$, as we have shown in the previous
section.

The spin-polarized regime is shifted slightly by increasing V,
due to the scattering via the positive
potential V encountered by the incident electrons from the leads attached to the corners of the triangle.
Potential V can be thought as a barrier that the electrons need to overcome, by acquiring sufficiently large incident energy E.
We have verified that the transmission probabilities $T^{\mu}_{ij}$
between different terminals satisfy the conditions $T^{\uparrow}_{ij}=T^{\downarrow}_{ji}$,
which are a consequence of the triangle's symmetry.
Combined with $T^{\uparrow}_{12}\approx 1,T^{\downarrow}_{12}\approx 0$ they lead to the conditions shown at the right side of Fig. \ref{fig1}.
As long as these conditions are satisfied the electrons propagating in opposite directions along
the edges of the flake, will have also opposite spins. In this sense, the electrons in the trigonal
Dirac flakes become helically spin-polarized along the edge
for sufficiently strong SOC. Additionally, we observe that the transmission is enchanched by the
SOC due to the accumulation of the wavefunction at the corners of the triangle where
the leads are attached, as can be seen in Fig. \ref{spectrum}d.
However the helicity supresses
the electronic flow for one of the spin orientations
and consequently the total (charge) transmission remains
almost one for every $\alpha$ at energies where the 
transmission occurs via the energy levels of the flake.

To further analyze the helical regime, in Fig. \ref{tsphase} we show a
density color-plot of the spin transmission $T^{s}_{12}$ versus E and $\alpha$,
along with individual cases for constant $\alpha$ in the bottom.
The helical(yellow) regime can be clearly distinguished where $T^{s}_{12}$
becomes maximum ($T^{s}_{12}=1$), signifying fully spin-polarized electrons
that propagate helically along the flake's edge. We notice that for larger V,
corresponding to materials with wider gaps, larger SOC
is needed in order to observe the helically polarized electrons.
This is due to the interplay between the gap generated by the staggered
potential and the one due to the SOC\cite{Kane}.
In order for the QSH effect to manifest, being essentially responsible
for the helicity, the SOC has to dominate the gap over the staggered potential.
The helical regime is retained for all flake sizes, ranging from tenth to a few hundred atoms, 
as can be seen in Fig. \ref{tssize}, with more fluctuations created, due to the denser energy levels
as the size is increased.

By appropriately tuning the chemical potentials $\mu_i$ corresponding
to each terminal i in the three-terminal setup\cite{Shen}, spin-polarized currents with opposite polarization
flow along the opposite sides of the triangle.
For $\mu_1=\mu_3$ spin currents $I^{s}_1=-I^{s}_3=\frac{e}{4\pi}(\mu_1 - \mu_2)$ flow at terminals 1 and 3
with opposite spin polarizations. Therefore, spin-resolved transport along the
different sides of the flake is possible,
allowing in this sense a geometrical manipulation of the electron spin,
since the path of the spin-polarized electrons can
be controlled by the flake's shape.

We expect similar phenomena for differently shaped
flakes such as hexagonal, which could offer additional possibilities
to control the spin dependent transmission by attaching more
leads at the flake's corners. Also impurities or
weak disorder should not affect our results since
the system is protected from backscattering due to the
time reversal symmetry, as in QSH systems.

\begin{figure}
\begin{center}$
\begin{array}{cc}

\includegraphics[width=0.3\columnwidth,clip=true]{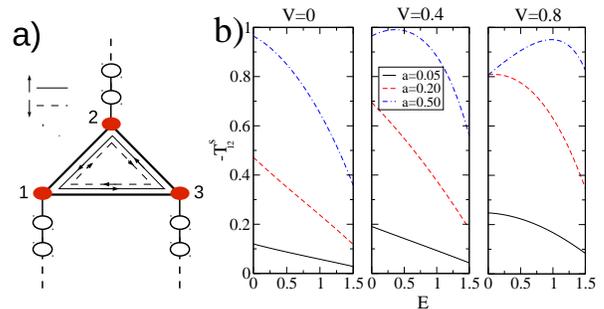}
\includegraphics[width=0.6\columnwidth,clip=true]{fig6b.eps}

\end{array}$
\caption{ a) A simple model simulating the three-terminal setup
of the trigonal Dirac flake, the edge state mechanism along
with the influence of the intrinsic SOC. b) $T^{s}_{12}$ for different $\alpha$ and V.
The curves describe the trend of the numerical results for all cases.}
\label{simple_model}
\end{center}
\end{figure}

In order to further investigate the helical mechanism
we introduce a simple model consisting of a triangle
with three sites with constant on-site potential V in analogy
with the potential at the corners of the flake in Fig. \ref{fig1}
and three linear chains attached to it's corners
acting as leads, as shown in Fig. \ref{simple_model}a.
Also, we consider spin dependent hoppings of strength $\alpha$ inside the triangle simulating
the SOC. This model can be considered as an isolation of the SOC
mechanism for each of the sublattices A or B (Eq. \ref{h_soc}) 
inside one hexagon in the honeycomb lattice, 
corresponding to L=1 for the trigonal
Dirac flake. Alternatively it can be thought as a
robust simulation of the numerical model, since it contains the
basic ingredients of the problem, namely the fact that the electrons propagate in
a straight orbit along the border of the trigonal flake due to the edge states,
under the influence of the intrinsic SOC.
We note that due to the way we oriented the triangle
the transmission from site 1 to 2 $T^{s}_{12}$ takes negative values.
In Fig. \ref{simple_model}b we show the spin transmission $T^{s}_{12}$
for different V and SOC strengths. We show 
 $-T^{s}_{12}$ in order to lie in the positive axis. The calculation details are presented in the Appendix.
The analytical curves catch the trend of the numerical results for the trigonal flake
with maximum $T^{s}_{12}\approx1$ for sufficiently large $\alpha$, which
is shifted across the energy for finite V.

\section{Concluding Remarks}

We have presented a numerical study of the spin properties
of electrons in flakes made of Dirac materials at the presence
of intrinsic spin-orbit-coupling (SOC).
A detailed analysis of the edge states 
is also presented, based on sublattice symmetry arguments.
The flakes are simulated by an effective tight-binding model,
which can be thought as a numerical version of the Dirac equation.
We probe the spin dependent transport by forming a multi-terminal device
which allows the calculation of the transmission probabilities
along the edges of the flakes via the Green's function formalism.

We have shown that the electrons propagating
along the border of the flakes via the edge states, become helically spin
polarized for strong SOC, for both the massless and massive
cases, corresponding to Dirac materials with and without a gap at the Fermi
energy, respectively. The helicity can be easily detected in a three-terminal
setup formed naturally by attaching 1d leads at the 
corners of a trigonal flake. Appropriate tuning of the chemical potential
of each lead, creates spin-resolved transport with opposite polarization between
the opposite sides of the flakes, allowing a geometrical manipulation
of the electron spin.

In other words, we have shown that spin-resolved transport can
be created intrinsically in flakes made of Dirac materials
by forming multi-terminal devices without the need of
external fields, the only requirement being sufficiently strong
intrinsic SOC. We hope that our work will motivate further 
investigation of the SOC effects in 2d materials in conjunction
with topological effects due to confinement in flakes and
other nano-structures.

\section{Acknowledgements}

We are grateful to Chi-Shung Tang, Shun-Jen Cheng and Victor A. Gopar for valuable discussions
and carefull reading of the manuscript. This work 
was supported by Ministry of Science and Technology, Taiwan
through No.\ MOST 103-2112-M-239-001-MY3, 
MOST-103-2112-M-009-014-MY3, and National Science
Council through No.\ NSC 102-2112-M-009-009-MY2.
Also, we thank the Ministry of Education through the Aiming for Top
University Plan (MOE ATU), and the National Center for Theoretical Sciences.

\subsection{Appendix: Simple model}

In the following Appendix, we present the derivation of the transmission probabilities 
for a simple system consisting of a triangle with three sites, and three linear 
chains attached to it's corners acting as leads. We consider SOC inside the triangle 
by assuming spin-flip hopping of strength $\alpha$ between each site, denoting 
also the SOC strength in analogy with the numerical model. 
Spin-up(down) corresponds to $\mu=1(-1)$.

The Hamiltonian of the system can be split in spin-up and spin-down
blocks since the SOC does not mix the two spins, as
\begin{equation}
H= \left( \begin{array}{cc}h^{\uparrow} & 0 \\ 0 & h^{\downarrow}
 \end{array} \right)
 \label{hamiltonian}
\end{equation}
where $h^{\uparrow}$ and $h^{\downarrow}$ are,
\begin{equation}
h^{\mu}= \left( \begin{array}{ccc}V & 1 - i\mu\alpha  & 1 + i\mu\alpha \\
1 + i\mu\alpha  & V & 1 - i\mu\alpha \\  1 - i\mu\alpha & 1 +i\mu\alpha & V
 \end{array} \right),
  \label{h_up}
\end{equation}
by assuming real hopping t=1~eV between the three sites, with $\mu=1(-1)$ for spin up(down).
The transmission probabilities between the different sites of the triangle
can by calculated via the Green's function
\begin{equation}
    G(E)=(EI-H-\Sigma(E))^{-1}
    \label{green}
\end{equation}
where $\Sigma(E)= \Sigma_{1}(E) + \Sigma_{2}(E) +  \Sigma_{3}(E)$
is the self-energy due to the semi-infinite linear chains
at the corners described by
$H = \sum_{i} c_{i}c^{\dag}_{i+1} +c.c$ with energy dispersion $E=2tcos(k)$.
$\Sigma(E)$ is given by equal diagonal blocks for spin up and down
\begin{equation}
\Sigma^{\mu}(E)= \left( \begin{array}{ccc}\Sigma_{1d}(E) & 0  & 0) \\
0  & \Sigma_{1d}(E) & 0 \\  0 & 0 & \Sigma_{1d}(E)
 \end{array} \right)
 \label{self_up}
\end{equation}
where $\Sigma_{1d}(E)$ is the self-energy of each chain,
which can be calculated recursively as follows \cite{Lewenkopf}.

In general we can calculate the Green's function via
\begin{equation}
G(E)=(E-g(E))^{-1}.
\end{equation}
where $g(E)$ represents the surface Green's function of the semi-infinite chain.
Adding a single site via hopping t=1 to the 
chain should not alter it's Green's function,
so that $G(E)=g(E)$. Consequently we can write the following recursive relation for $g(E)$,
\begin{equation}
g(E)=(E-g(E))^{-1} \Rightarrow g(E)=\frac{E}{2} \pm i\sqrt{1-\frac{E^{2}}{4}}.
\end{equation}
Since we are interesting only in the retarded Green's function
we choose the minus sign representing outcoming waves from the point of excitation,
giving the self-energy $\Sigma_{1d}(E)=g(E)$,
\begin{equation}
\Sigma_{1d}(E)=\frac{E}{2} - i\sqrt{1-\frac{E^{2}}{4}}  \text{ for } -2<E<2.
\end{equation}
Since H and $\Sigma(E)$ are both block-diagonal in the spin basis, this is also true for Eq. \ref{green}.
Therefore the Green's function elements between opposite spins are zero, and we can write
\begin{equation}
    G^{\mu}(E)=(EI-h^{\mu}-\Sigma^{\mu}(E))^{-1}
    \label{green_up}
\end{equation}
By plugging Eq. \ref{h_up}, Eq. \ref{self_up} in Eq. \ref{green_up} and inverting,
we derive the following formula for the Green's function element from site 1(excitation)
to 2(response)
\begin{widetext}
 \[
G^{\mu}_{21}(E) = -\frac{1-\alpha^{2}+E-\Sigma_{1d}(E)+
i\mu\alpha(-2+E-\Sigma_{1d}(E)-V)-V}{(2-E+\Sigma_{1d}(E)+V)
(-3\alpha^{2}+(-1-E+\Sigma_{1d}(E)+V)^{2})}
\]
\end{widetext}
In order to calculate the respective transmission probabilities
we apply the Fischer-Lee relations,
\begin{equation}
    T_{12}^{\mu}(E)=v(E)^{2}|G_{21}^{\mu}(E)|^{2}
    \label{t_up}
\end{equation}
where $v(E)=\frac{\partial E}{\partial k}=-2\sqrt{1-\frac{E^{2}}{4}}$
 is the group velocity of the linear chains,
assuming that $\hbar=1$.
We note that the transmission probabilities between opposite spins are zero
since the respective Green's function elements are zero.

By using the above relation we can calculate the spin transmission
 $T_{12}^{s}(E)$ which is plotted in Fig. \ref{simple_model}
\begin{equation}
    T_{12}^{s}(E)= \left(T_{12}^{\uparrow}(E) - T_{12}^{\downarrow}(E) \right).
    \label{Gs}
\end{equation}

\end{document}